\documentstyle[seceq,graphicx]{ptptex}
\title{%        %You can use \\ for explicit line-break
Compton thick AGN in the {\it Suzaku} era
}
\author{%       %Use \sc for the family name
Andrea {\sc Comastri}$^1$, Roberto {\sc Gilli}$^1$, 
Cristian {\sc Vignali}$^2$, Giorgio {\sc Matt}$^3$, Fabrizio {\sc Fiore}$^4$, 
Kazushi {\sc Iwasawa}$^5$
}

\inst{%         %Affiliation, neglected when [addenda] or [errata]
$^1$ INAF -- Osservatorio Astronomico di Bologna, Bologna, Italy 
\\
$^2$  Dipartimento di Astronomia, Universit\`a di Bologna, Bologna, 
Italy  
\\
$^3$ Dipartimento di Fisica, Universit\`a di Roma Tre, Roma, Italy  
\\
$^4$ INAF -- Osservatorio Astronomico di Roma, Monteporzio, Italy
\\
$^5$ MPE --  Garching, Germany}

\recdate{%      %Editorial Office will fill in this.
}

\abst{%       %this abstract is neglected when [addenda] or [errata]
{\it Suzaku} observations of two hard X--ray ($>$ 10 keV) selected 
nearby Seyfert 2 galaxies are presented.
Both sources were clearly detected with the {\tt pin} Hard X--ray Detector 
up to several tens of keV, allowing for a fairly good characterization of the broad 
band X--ray continuum. Both sources are heavily obscured, one 
of which ({\tt NGC 5728}) being Compton thick, while at lower energies the shape and 
intensity of the scattered/reflected continuum is very different. Strong iron K$\alpha$ 
lines are detected in both sources. There are also hints for the presence 
of a broad relativistic iron line in {\tt NGC 4992}.
}

\begin{document}

\maketitle

\section{Introduction}
A fraction as high as 50\% of Seyfert 2 galaxies in the nearby Universe are
obscured in the X--ray band by column densities of the order of, 
or larger than the inverse of the Thomson cross-section 
($N_H\ge \sigma_T^{-1} \simeq 1.5 \times 10^{24}$~cm$^{-2}$),  
hence dubbed Compton thick (CT). If the optical depth ($\tau = N_H \sigma_T$) 
for Compton scattering does not exceed values of the order
of ``{\it a few}", X--ray photons with energies higher than 10--15 keV are able 
to penetrate  the obscuring material and reach the observer. For higher values
of $\tau$, the entire X--ray spectrum is depressed by Compton down scattering 
and the X--ray photons are effectively trapped by the obscuring material 
irrespective of their energy. The former class of sources (mildly CT) can be efficiently 
detected by X--ray instruments sensitive above 10 keV, while for the 
latter (heavily CT) their nature may be inferred  
through indirect arguments, such as the presence of a strong iron 
K$\alpha$ line over a flat reflected continuum. 
The search for and the characterization of the physical 
properties of CT AGN is relevant to understand the 
evolution of accreting Supermassive Black Holes (SMBHs). In particular,  
mildly CT AGN are the most promising candidates to explain the 
residual (i.e. not yet resolved) spectrum of the X--ray background 
around its 30 keV peak (Comastri 2004a; Worsley et al. 2005) 
but only a handful of them are known beyond the local
Universe (see Comastri 2004b for a review). If this were the case, we may 
be missing a not negligible fraction of the accretion power in the Universe and
of the baryonic matter locked in SMBH (Marconi et al. 2004).
An unbiased census of extremely obscured AGN would require to survey the hard 
X--ray sky above 10 keV with good sensitivity. Such an argument is one of the
key scientific drivers of the {\sc SimbolX} mission (Ferrando et al. 2006), 
which will be hopefully launched in the next decade.
For the time being one has to rely on the observations obtained by 
the high energy detectors on board {\sc BeppoSAX}, {\sc INTEGRAL}, {\sc Swift}
and, more recently, {\it Suzaku}. Though limited to bright and thus low
redshift sources, they have proven to be quite successful in finding 
heavily obscured CT AGN.
As a first step forward towards a census of CT AGN we have conceived
a program with {\it Suzaku} to observe hard X--ray selected
bright AGN from the {\sc INTEGRAL/IBIS} (Beckmann et al. 2006) and {\sc Swift/BAT} 
(Markwardt et al. 2005) catalogues. 
The goal of this program is to discover {\it ``new"} CT AGN which are likely
to be present among the already detected sources, but not recognized 
as such due to the poor counting statistics and/or the lack of information
at lower energies. 
In order to select the most suitable candidates, we have considered the 
sources in the above mentioned AGN catalogues with a bright 
hard X--ray flux and tentative evidence of intrinsic absorption from observations at lower energies.
For a few of them the column densities are estimated to be close to the
CT threshold.  {\it Suzaku} observations were obtained for
NGC 5728 and NGC 4992.

\section{The Suzaku observations}

The reprocessed (v1.2) data were reduced using standard calibration
products available in November 2006. Source spectra are obtained from the
Front Illuminated  XIS chips with an extraction radius of $\sim$3$^{\prime}$, while background 
spectra are extracted from nearby regions with a larger radius to guarantee 
good statistics.  
The effective exposure time for both sources is of the order of 30 ksec.
The {\tt pin} hard X--ray source spectra were obtained taking into account both 
the instrumental background appropriate for each observation and the cosmic X--ray 
background. The {\tt pin}/XIS intercalibration constant was fixed at 1.16.
In the following, we report the basic results obtained from the analysis of the 
X--ray spectra of the two sources and refer to Comastri et al. (2007, in preparation) 
for a more exhaustive description of the data analysis and interpretation.

%%%%%%%%%%%%%%%%%%
\begin{figure}
\includegraphics[angle=-90,width=0.48\textwidth]{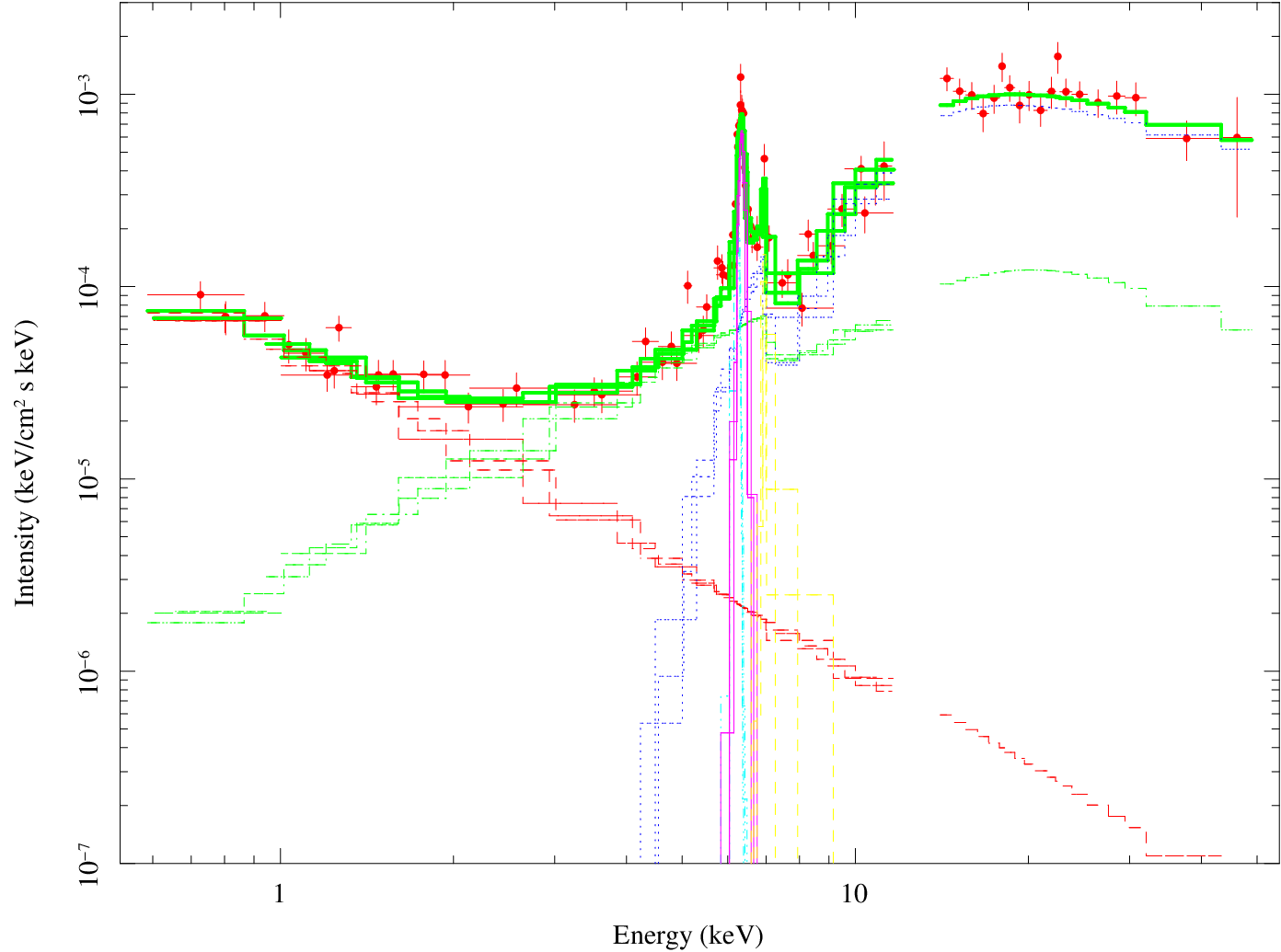}
\hfill
\includegraphics[angle=-90,width=0.48\textwidth]{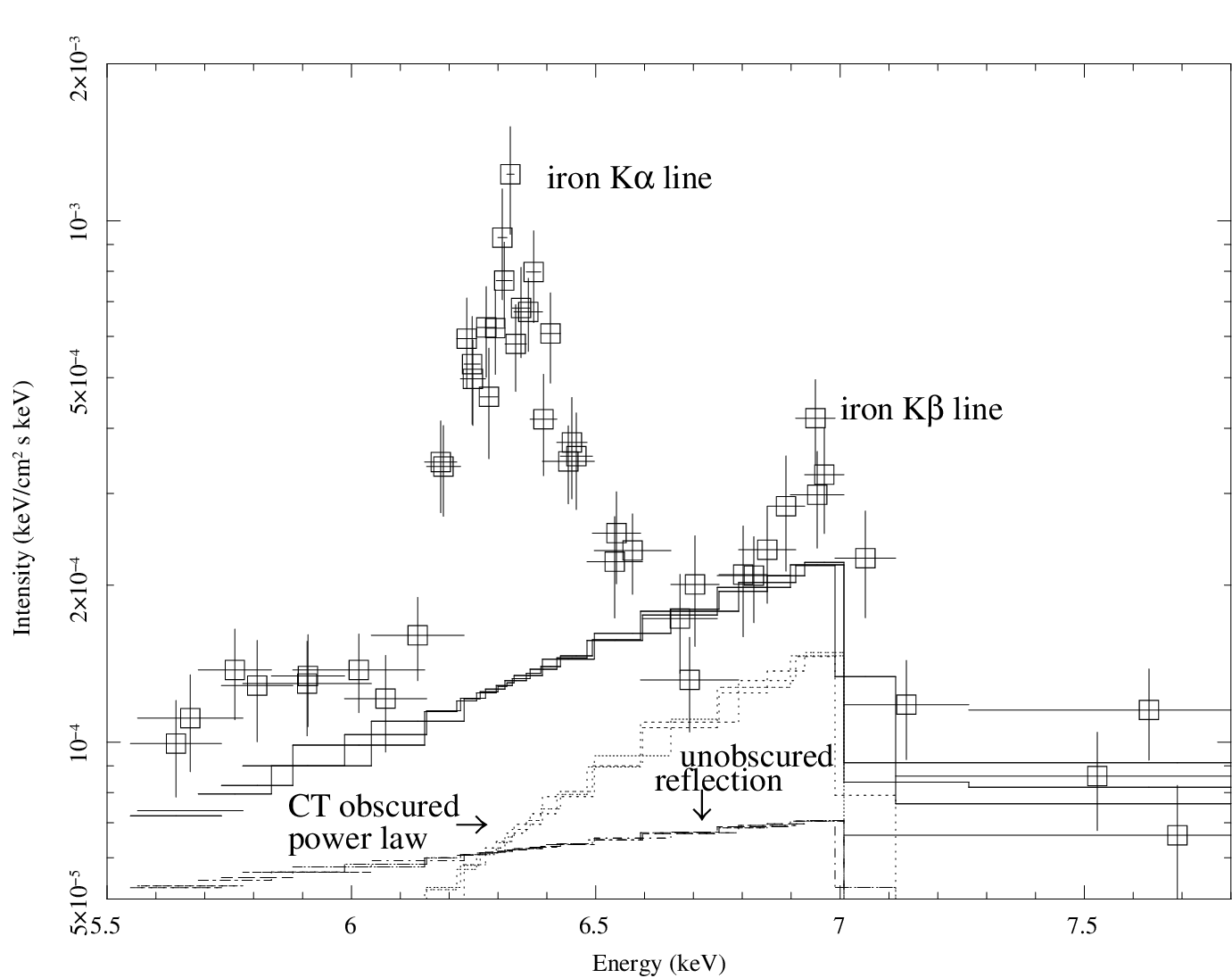}
\caption{
{\it Left panel}: The unfolded broad band  spectrum of NGC 5728 with the various components 
used to model the continuum and the iron lines. {\it Right panel}:  
A zoom on the "iron band" showing a strong K$\alpha$ line at $\sim$ 6.4 keV 
and a less prominent $K\beta$ line ($\sim$ 7 keV)  
on top of the underlying continuum (upper line) made by the sum of a CT obscured power law 
(middle line) and an unobscured reflected component (lower line).
}
\label{ngc5728}
\end{figure}
%%%%%%%%%%%%%%%%%%
%%%%%%%%%%%%%%%%%%
\begin{figure}
\includegraphics[angle=-90,width=0.48\textwidth]{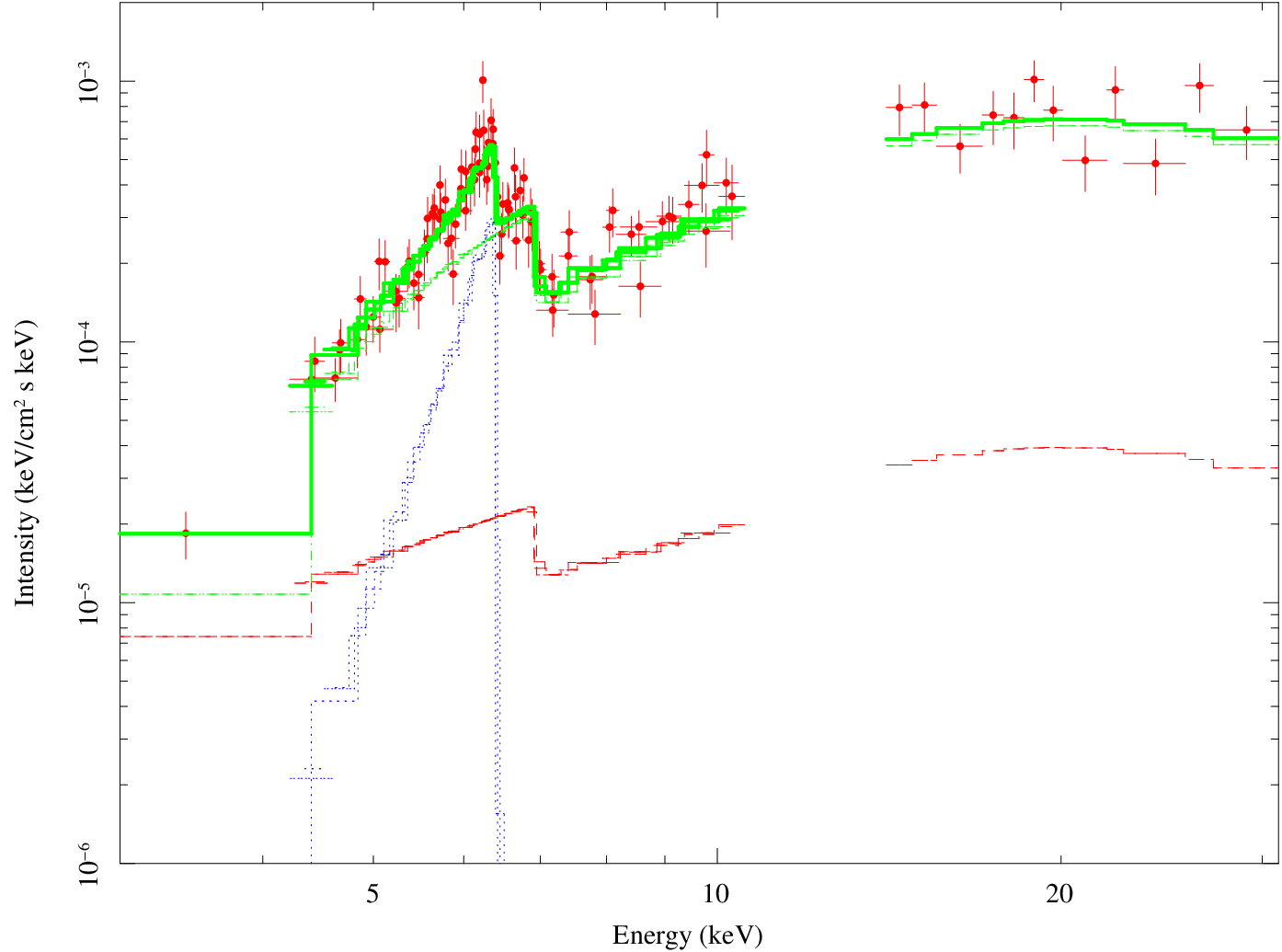}
\hfill
\includegraphics[angle=-90,width=0.48\textwidth]{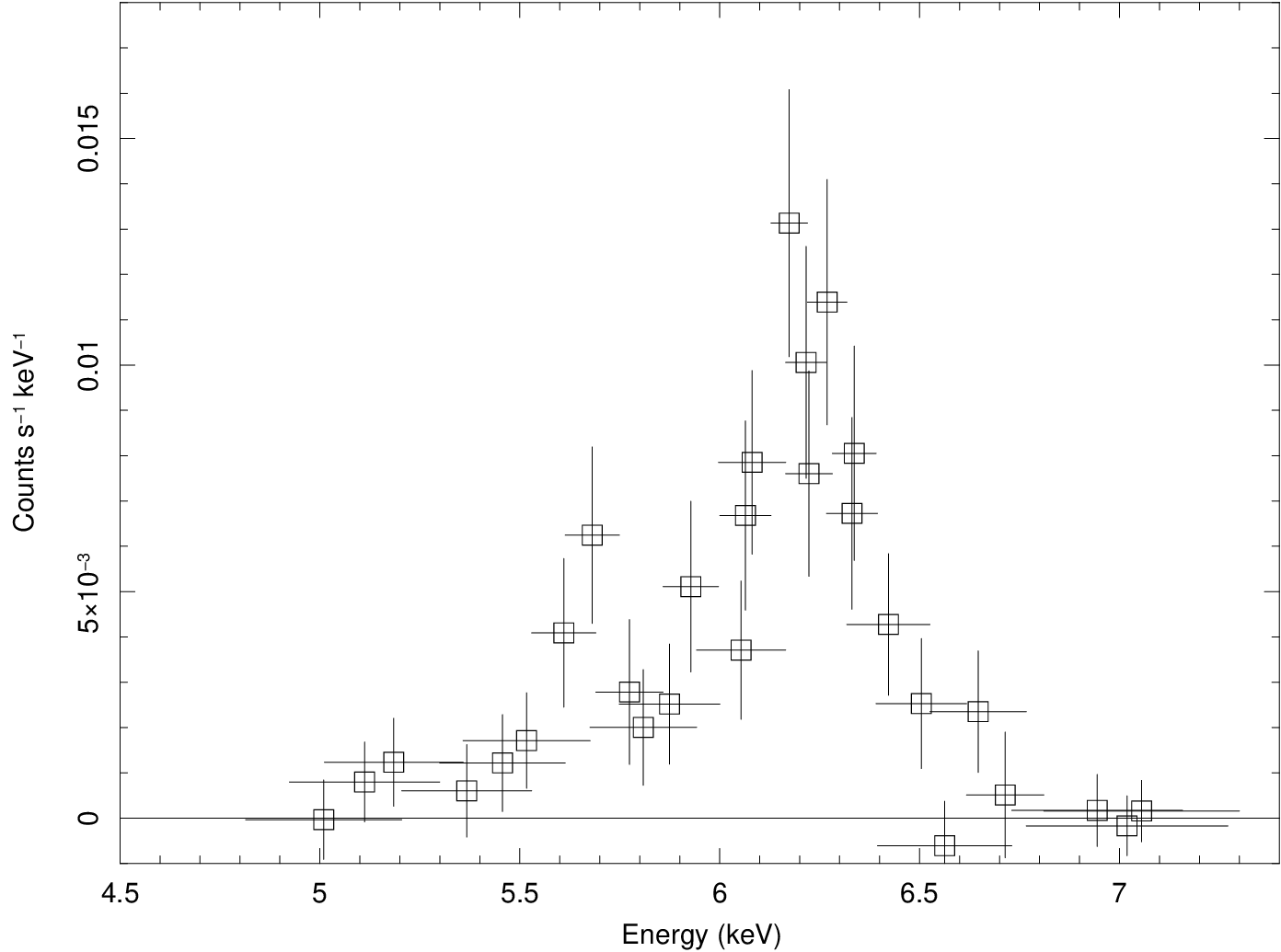}
\caption{
{\it Left panel}: The unfolded broad band spectrum of NGC 4992. The reflection dominated 
absorbed continuum, the relativistic disk line and a weak unobscured reflected component
are reported. {\it Right panel}: The residuals vs. the best fit continuum in the 4.5--7.5 
keV band. 
}
\label{ngc4992}
\end{figure}
%%%%%%%%%%%%%%%%%%

\subsection{NGC 5728}

The {\it Suzaku} spectrum of NGC 5728 is shown in Fig.~1 (left panel). 
The source is clearly detected 
by the {\sc pin} detector up to about 50 keV. The primary X--ray continuum 
is absorbed by Compton thick gas ($N_H \simeq 2.1\pm 0.2 \times 10^{24}$ cm$^{-2}$).
The power law slope has been fixed at $\Gamma$ = 1.9 due to the narrow energy 
range (20--40 keV) over which the continuum is free from obscuration effects.
At lower energy, the continuum can be represented by a two component model:
a flat one responsible for most of the X--ray flux in the $\sim$ 2--6 keV energy range 
and a steep one taking over below 2 keV. The former may be ascribed to reflection 
of cold material presumably from the inner wall of the torus, while 
the latter has a power law shape and can be identified as primary emission scattered 
by off--nuclear gas into the line of sight, or unresolved soft X--ray emission lines, 
as commonly observed in Seyfert 2 galaxies (Guainazzi \& Bianchi 2007).
The scattered/reflected flux accounts for  1--2 \% of the total unabsorbed flux
($\sim 5 \times 10^{-11}$ erg s$^{-1}$ cm$^{-2}$) in the 2--10 keV band.
The 2--50 keV unabsorbed luminosity is 2.3 $\times$ 10$^{43}$ erg $^{-1}$, 
typical of a bright Seyfert galaxy. It is interesting to note that the hard ($>$ 10 keV) 
X--ray flux as measured by the {\tt pin} detector is consistent within 20\% 
with the {\sc Swift/BAT} measurement in the overlapping energy range. 
A zoom of the $\sim$ 5--8 keV unfolded spectrum is shown in Fig.~1 (right panel). 
The iron line 
complex is best fitted with two gaussian lines: a strong (EW $\simeq 1.0 \pm 0.3$ keV) iron 
K$\alpha$ line at $\sim$ 6.4 keV and a K$\beta$ (EW $\sim 130 \pm 70$ eV) at $\sim$ 7 keV.
The relative ratio is consistent with that expected from cold neutral gas. 
The addition of a Compton shoulder parameterized by a Gaussian profile centered at 
6.3 keV and $\sigma$ = 40 eV (Matt 2002), though not statistically required, accounts 
for some 10\%  of the K$\alpha$ line flux, in reasonably good agreement with the 
value expected for reflection from Compton thick matter.

\subsection{NGC 4992}

The Seyfert 2 galaxy NGC 4992 is detected by {\it Suzaku} up to about 30 keV with a flux
consistent (within 10\%)  with that reported by {\sc INTEGRAL}. 
The continuum (a power law with $\Gamma$=1.9) 
is heavily obscured ($N_H \sim 4.5 \pm 0.5 \times 10^{23}$ cm$^{-2}$) 
but not Compton thick. The high energy spectrum is best fitted by adding a strong, absorbed, 
disk reflection component to the primary power law. The quality of the data is not such to tightly 
constrain the intensity of the reflection component. The 90\% lower limit ($R >$ 5) indicate 
a reflection dominated spectrum which is similar to that reported by Miniutti et al. (2007) 
from the analysis of the {\sc XMM-Newton} data of {\sc IRAS 13197-1627}. The source is extremely weak 
below 3--4 keV. The addition of an unabsorbed reflection spectrum only marginally improves the 
fit (Fig.~2, left panel). A zoom of the residuals in the 4.5--7.5 keV range, 
wrt the best fit continuum model, is shown in Fig.~2 (right panel). 
The shape of the residuals suggests the presence of a broad line. Indeed 
the best fit to the line emission is obtained with a diskline model. 
Leaving only the line flux and the
disk inclination angle as free parameters, the line equivalent width is $\sim 750 \pm 200$ eV 
and the inclination angle is $<$ 40 degrees (at 90\% confidence).  
The best fit EW is consistent with a reflection dominated nature of the broad band spectrum. 
The absorption corrected 2--50 keV luminosity is $\sim 6 \times 10^{43}$ erg s$^{-1}$.

\subsection{Epilogue} 

Relatively shallow {\it Suzaku} observations of two hard X--ray selected (with 
INTEGRAL/IBIS and Swift/BAT) nearby Seyfert 2 galaxies have revelead a wealth of
spectral complexity in their X--ray spectra. The good sensitivity over a broad X--ray energy range
makes {\it Suzaku} very efficient to study the most obscured sources in the 
nearby Universe and will eventually allow us to establish the AGN absorption distribution
at high column densities.

\section*{Acknowledgements}
We thank G. Miniutti for extremely useful discussions. 
Support from the Italian Space Agency (ASI) under the 
contract ASI-INAF I/023/05/0 is acknowledged.

\end{document}